\documentstyle[aps,preprint]{revtex}

\global\parskip = 2pt plus 0.2pt

\def\endignore{}
\def\ignore #1\endignore{}

\ifx\epsffile\undefined
\def\insertfig#1{}
\else
\def\insertfig#1{{\baselineskip=4pt
\centerline{\epsfxsize=\hsize\epsffile{#1}}}}\fi

\def\beq{\begin{equation}}
\def\eeq{\end{equation}}

\def\beqa{\begin{eqnarray}}
\def\eeqa{\end{eqnarray}}

\def\Dsl{\hbox{\kern.1em/\kern-.7000em$D$}} 

\def\mybar#1{\kern 0.8pt\overline{\kern -0.8pt#1\kern -0.8pt}\kern 0.8pt}
\def\sla#1{\raise.15ex\hbox{$/$}\kern-.57em #1}
\def\Sla#1{\kern.15em\raise.15ex\hbox{$/$}\kern-.72em #1}

\def\roughly#1{\mathrel{\raise.3ex\hbox{$#1$\kern-.75em%
    \lower1ex\hbox{$\sim$}}}}

\def\Re{\mathop{\rm Re}}

\def\ChPT{\raise.45ex\hbox{$\chi$}PT}

\newcommand{\emph}[1]{{\em #1}}

\newcommand{\Group}[2]{\hbox{{\sl #1}($#2$)}}
\newcommand{\U}[1]{\Group{U\kern0.05em}{#1}}
\newcommand{\SU}[1]{\Group{SU\kern0.1em}{#1}}
\newcommand{\SL}[1]{\Group{SL\kern0.05em}{#1}}
\newcommand{\Sp}[1]{\Group{Sp\kern0.05em}{#1}}
\newcommand{\SO}[1]{\Group{SO\kern0.1em}{#1}}

\begin{document}
\tighten
\preprint{\vbox{
\hbox{HUTP-95/A044}
\hbox{hep-th/9511123}}}
\title{Dynamical Determination of Dilaton and Moduli\\
Vacuum Expectation Values}

\author{Rulin Xiu%
\footnote{Address starting January 1996:
Department of Physics, University of Maryland, College Park MD 20742.
E-mail: {\tt xiu@huhepl.harvard.edu}}
\medskip}

\address{Lyman Laboratory Of Physics \\
Harvard University\\
Cambridge, MA 02318\medskip}

\date{November 1995}

\maketitle

\begin{abstract}
We determine the dilaton and moduli vacuum expectation values using the
one-loop effective potential and topological constraints.
A new ingredient of this analysis is that we use a dilaton K\"ahler
potential that includes renormalization effects to all loops.
We find that the dilaton vacuum expectation value is related to certain
topological properties of the compact spacetime.
We demonstrate that values of the dilaton vacuum expectation value that
are consistent with the weak scale measurements can be  dynamically
obtained in this fashion.
\end{abstract}


The determination of the dilaton ($S$) and moduli vacuum expectation
values (VEV's) is an important problem in  string phenomenology
because it is directly related to the predictions of the string models
\cite{wdr}.
The dilaton and moduli fields are flat directions to all orders
in string perturbation theory, and it is hoped that some non-perturbative
effects will lift these flat directions and determine these VEV's
dynamically.

Non-perturbative gaugino condensation in the hidden sector is a ready
candidate for specifying the dilaton and moduli VEV's.
The hidden sector of the heterotic string is asymptotically free and
will ``confine'' at a low energy and give rise to a dilaton-dependent
superpotential.
However, it is found that this superpotential generically leads to a
``runaway'' dilaton VEV, {\em i.e.}\ the potential energy is minimized
at either $S \rightarrow 0$ or $S \rightarrow \infty$.

So far, there are two types of plausible string-inspired gaugino
condensation scenarios.
One is the $c$-number scheme \cite{wit2,iu} in which it is assumed that a
constant term is induced in the superpotential by the gaugino condensation
to cancel the dilaton dependent contribution.
The other is multiple gaugino condensation \cite{kay,dixon,tom3,cas}, in
which multiple gauge groups condense in the hidden sector.

In this work, we discuss the determination of the dilaton
and moduli VEV's through the one-loop effective potential of
the gaugino condensation dynamics in the hidden sector.
We will discuss particularly the
$c$-number gaugino condensation models proposed in \cite{iu}
for string models with  no-scale structure and in which the
gauge coupling constant does not receive string threshold
corrections.  The no-scale structure plays the crucial role
in generating the large mass hierarchy in these models.
Because of the no-scale structure, the dialton and moduli VEVs are not
determined at the tree level and  the examination of  the  one-loop
effective potential becomes necessary.

The other purpose of this paper is to examine  whether the inclusion
of  higher-loop corrections to the dilaton K\"ahler potential
improves the dilaton runaway problem.
In ref.~\cite{loop1}, we pointed out that the one-loop corrections
to the dilaton K\"ahler potential may play a significant role in the
gaugino condensation dynamics and in solving the dilaton runaway problem.
Because the gauge coupling constant is closely related to the dilaton
K\"ahler potential, and the gauge coupling is large at the gaugino
condensation scale, the inclusion of higher loop correction becomes
very important, which is partly the subject of this work.

The new ingredient of the present work is that we use a dilaton
K\"ahler potential derived in Ref.~\cite{all}, in which
renormalization effects to all loops are taken into account.
In our approach, both the effective potential and the quantization
condition imposed by the compact manifold geometry or topology
play important roles in determining the dilaton and moduli VEV's.

In Ref.~\cite{all} the $E_8$ gauge coupling and the modified
dilaton--axion K\"ahler potential are derived to all orders for
effective field theories arising from 4-dimensional $N = 1$ heterotic
string models in which the gauge coupling constant does not receive string
threshold corrections.
In this paper, we will use the formulation of Ref.~\cite{all} and
analyze its one-loop effective potential using the same methods as in
Ref.~\cite{loop1}.

We first describe the model we will analyze.
The K\"ahler potential given in Ref ~\cite{all} is
\beq
K = \log \left[\frac{1}{2}g^2\left(1+\frac{b_0}{3}g^2\right)^{-3}\right].
\eeq
Here $g^2$ is the gauge coupling constant, which is related to the
dilaton and the effective gaugino bilinear field $H$ through
\beqa
g^{-2} & = & \frac{1}{2}(S + \bar{S})
- \frac{2b_0}{3} \log g^2 + b_0 \log |H|^2
+ \frac{1}{2}a_0,
\\
a_0 & \equiv & b_0 \left(-2 + \frac{10}{3} \log 2\right).
\eeqa
We assume the usual superpotential obtained by symmetry arguments
\beq
W = d \left[ \frac{1}{4} S H^3 + \frac{b_0}{2} H^3
\log(\eta H) \right] + c_0 + W_0.
\eeq
Here $c_0$ and $W_0$ are the contributions from the charged background
VEV's and matter fields, respectively.
The parameters $d$ and $\eta$ are not fixed by symmetry requirements,
but are specified by the underlying gaugino condensation dynamics.
In Ref ~\cite{glo,lo} it was shown that $d=1$, and we will use this result
in the analysis below.
We follow the argument in Ref ~\cite{loop1} and treat the $H$ field as a
dynamical field.
Additional details of the calculation is given in
Refs.~\cite{mary3,loop1}

In the above model, the inverse of  K\"ahler matric is
\beq
(K^{-1})^{i\bar{j}}=\pmatrix{1/x_2 + 4b_0^2C / (3|H|^2) &
-2b_0C / (3H) & -\frac{2}{3} b_0C \cr
         -2b_0C/{3\bar{H}} & \frac{1}{3} {C} & \frac{1}{3} {HC} \cr
         -\frac{2}{3} {b_0C}& \frac{1}{3}{\bar{H}C}& \frac{1}{3}C(|H|^2+C)
\cr},
\eeq
where
\beq
 C\equiv T+\bar{T}-|H|^2,
\eeq
and
\beqa
x_2 & \equiv & \left. \frac{dx_1}{dy} \right/ \frac{df}{dy}
= \frac{y}{4(y+b_0/3)^2(y-2b_0/3)},
\\
x_1 & \equiv & \left. \frac{dL}{dy} \right/ \frac{df}{dy}
= -\frac{1}{2(y+b_0/3)},
\\
L &\equiv & \frac{1}{2} g^2 \left( 1+\frac{b_0}{3}g^2 \right)^{-3}
\nonumber\\
& = & \frac{1}{2} y^2 \left( y+\frac{b_0}{3} \right)^{-3},
\\
f & \equiv & 2y +\frac{4b_0}{3}\log y + \frac{1}{2}c_0
\nonumber\\
& = & S + \bar{S} b_0 \log |H|^2,
\\
y & \equiv & 1/g^2.
\eeqa
Here $g^2$ is the gauge coupling constant at the gaugino condensation
scale, which is determined by the parameter $\eta$ through the eq.(2).
In the following analysis, we will express everything in terms
of $g^2$ instead of $\eta$.

Solving the tree-level vacuum condition
\beqa
\label{tvacc}
\langle \tilde{W}_H \rangle \equiv \left\langle
W_H - \frac{L_H}{L} W \right\rangle = 0,
\\
\label{tvaccc}
\langle \tilde{W}_S \rangle \equiv \left\langle]
W_S + K_S W \right\rangle = 0,
\eeqa
we obtain
\beqa
H & = & \frac{1}{\eta} e^{-S / (2b_0)},
\\
\langle W \rangle &=& c_0 = \frac{d}{4x_1} \langle H^3 \rangle.
\eeqa

Using the vacuum conditions eqs.~(\ref{tvacc}) and (\ref{tvaccc})
and assuming that $S=\bar{S},H=\bar{H}$ ({\em i.e.}\ $CP$ violation is
highly suppressed), we
obtain the normalized fermion mass eigenvalues
\beqa
m_{1,2}^{1/2}=e^G\left[ \frac{1}{2}(1-x)\pm
\sqrt{\frac{1}{4}(1-x)^2+3xz}\, \right],
\qquad m^{1/2}_3=0.
\eeqa
The scalar masses are
\beqa
M^{s2}_{1,2}&=&e^G\left[ \frac{1}{2}(2-x)^2+3xz\pm \frac{1}{2}
(2-x)\sqrt{(2-x)^2+3xz}\, \right],
\\
M^{s2}_{3,4}&=&e^G \left[ \frac{1}{2}x^2+3xz\pm \frac{1}{2}
x\sqrt{(x^2+3xz}\, \right], \\
M^{s2}_{5,6}&=& 0.
\eeqa
The gravitino mass is
\beq
M_{3/2}=e^G=\frac{1}{8}y^2(y+\frac{b_0}{3})^{-1}z^{-3}.
\eeq
In the above
\beq
x \equiv \frac{x_1^2}{x_2} = 1-\frac{2}{3}b_0g^2, \qquad
z \equiv  C/|H|^2.
\eeq
One can see that in the case of \cite{loop1},
 $x=1$, the above computation yields
the same result as ref ~\cite{loop1}.

As in the one-loop case \cite{loop1},
the above result indicates that  dilaton mass is on the same order
as the $H$ mass (to be identified
with the scale of gaugino condensation) {\em independently of the
value of the dilaton VEV}.
Supersymmetry is broken in the hidden sector.
Furthermore,
the all-loop effect split the degeneracies between
the dilaton and the $H$ field mass in the one-loop case.

Next we examine the vacuum structure using the one-loop effective
potential.
The potential energy only depends on the modular
invariant function $z = {C}/{|H|^2}$, and therefore the moduli
and dilaton VEV's are not uniquely determined by
the effective potential in this model.
We take the cut-off scale to be
\beqa
\Lambda^2 & = & |H|^2e^{K/3}.
\eeqa
The one-loop effective potential is
\beqa
v_1 &=& 64 \pi^2 V^{\rm 1-loop}=
2\, {\rm Str}(M^2\Lambda^2)+
{\rm Str}\left [ M^4\log(M^2/\Lambda^2)\right ].
\eeqa
Given a value for $y$, we can minimize the effective potential with respect
to $z$.  For $y \sim 1$ we find a global minimum at $z \sim 1$.
For example, for $E_8$ hidden sector $b_0=90/16\pi^2$,
$z=0.63 $ for $y=1 $,   $z=0.36$ for $y=0.1$.
For  the $E_6$ hidden sector, $b_0=36/16\pi^2$,
 $z=0.73$ for $y=1 $ and $z=0.22 $ for $y=0.1 $.

The above result does not yield the hierarchy between the
moduli VEV and the gaugino condensation
scale.  Therefore,  the proposal  in
\cite{loop1}  that  the inclusion of the higher loop correction
to the dilaton K\"ahler potential  might  generate a large $z$
 does not work here. In fact, one can see this immediately
from the K\"ahler potential in \cite{all}. It is proposed in \cite{loop1}
that  the dilaton K\"ahler potential getting large at the gaugino
condensation scale may lead to the large $z$.  However,
the all-loop corrected dilaton K\"ahler potential  in \cite{all}
does not become large as the gauge coupling constant gets large
at the gaugino condensation scale.

It is interesting to notice, however, that in this type of string models
$z = C/|H|^2\sim (T+\bar{T})e^{(S+\bar{S})/2b_0} \sim 1$
does not necessarily mean that dilaton VEV runs away
to $\langle S \rangle \rightarrow 0$.
We may still obtain dilaton VEV consistent with the weak scale measurement
if we allow the compactification radius to be comparable to the gaugino
condensation scale,
{\em i.e.}\ $T \sim R^2 \sim |H|^2 \sim e^{-(S+\bar{S})/2b_0}$.
In fact, in a dynamical symmetry breaking scheme proposed in Ref.~\cite{iu},
a compactification radius comparable to the gaugino condensation scale
is a requirement in most known string models.
In the following, we will briefly review these arguments, and then we will
discuss the determination of the dilaton and moduli VEV's in these models.

In Ref ~\cite{iu}, we proposed that gaugino condensation
may induce the charged background VEV's which breaks gauge symmetry.
This dynamics makes possible affine level one grand unified
({\em i.e.}\ $SU(5)$ or $SO(10)$) string models
with intermediate gauge symmetry breaking scale
$M_{\rm GUT} \sim 10^{16}$~GeV.
We show that, in some string models, the quantization
condition enforces the compactification radius to be on the order of
the gaugino condensation scale.
For example, in the orbifold models, the $Z_n$ symmetry requires the
Wilson lines to be quantized:
$\oint A^i dx_i = 2\pi / n$.
For an orbifold with the moduli  background fields
\beq
G^{ij} = R^2 \delta^{ij},
\qquad G_{ij}= R^{-2} \delta^{ij},
\eeq
one has
\beq
\oint A^i\cdot dx_i = \oint A^i G_{ij} dx^i
= \frac{2\pi A}{R} = \frac{2\pi}{n},
\qquad A = \frac{R}{n}.
\eeq
The compactification radius is related to the
charged background VEV which is of the order of the gaugino condensation
scale.

In the following, we will determine the dilaton and moduli VEV using
the above quantization condition and the value for $z=C/|H|^2$
which is dynamically determined through the one-loop effective
potential given above.
To do this, we use the relation between the compactification radius and
the real part of dilaton $\Re(S)$ and moduli VEV $\Re(T)$ \cite{ine,wdr}:
\beq
R^2 = \Re(S) \cdot \Re(T),
\eeq
we get
\beq
\Re(S) = \frac{n^2 A^2}{\Re(T)}.
\eeq
The $A$ is related to the $H$ through the tree-level vacuum condition
\beqa
c_0\equiv d_{ijk}A^iA^jA^k = -\frac{dH^3}{4 x_1}.
\eeqa
Assuming  that all the induced charged background is the same
{\em i.e.}\ $A_i = A$ and $d_{ijk} \sim 1$, we obtain
\beqa
A = \left( \frac{-1}{4x_1} \right)^{1/3} H.
\eeqa
We then obtain
\beqa
Re(S) & = & n^2 \left( \frac{1}{4x_1} \right)^{2/3}
\frac{H^2}{Re(T)}
\\
& = & n^2 \left[ \frac{1}{2} \left( y+\frac{b_0}{3} \right)
\right]^{2/3} \frac{2}{z+1}.
\eeqa
It is easy to see that  with the dynamically determined $z$ and $y$,
the dilaton VEV can be determined which also depends  on the integer
$n$ of $Z_n$ symmetry. For example, for $E_6$ hidden sector,
with $y=1$, $z=0.73$ and
$Re(S)=0.76n^2$; with $y=0.1$, $z=0.22$ and $Re(S)=0.32n^2$.
We see that realistic VEV's for dilaton can be easily obtained.
Although the calculation performed here is at best a model of
the full gaugino condensation dynamics, the above analysis
indicates  that  the specification of $z$ and
the weak scale gauge coupling constant measurement could uniquely
determine the $n$; on the other hand, with the known string model,
$Re(S)$ can be dynamically determined.

We conclude that the hierarchy between the moduli VEV and the
gaugino condensation scale is not generated with the inclusion
of the high-loop corrections to the dilaton K\"ahler potential.
However,   the  moduli and dilaton  VEV's,
consistent with weak scale measurements, can be dynamically
determined  in the  models proposed in Ref.~\cite{iu}  by using
 imposed  quantization conditions.
A more reliable calculation of dilaton and moduli VEV's in this scenario
depends on a better understanding of the gaugino condensation dynamics,
since the dilaton mass is on the order of the gaugino condensation scale.
We hope the recent  progress on the dual descriptions of the dynamics
may shed some light on this.

\section*{Acknowledgments}
We thank M. Gaillard for sharing  related work and helpful discussions.
I especially thank M. Luty for supporting this work.

\end{document}